\documentclass[12pt]{article}
\usepackage{times}
\usepackage{geometry}
\usepackage[utf8]{inputenc}
\usepackage{hyperref}
\usepackage{enumitem,amssymb}
\usepackage{amsmath}
\usepackage{aas_macros}
\usepackage{graphicx}
\usepackage{ragged2e}
\usepackage{natbib}
\setcitestyle{aysep={}} 
\newlist{thematic}{itemize}{8}
\setlist[thematic]{label=$\square$}
\usepackage{pifont}

\usepackage{titlesec}
\titleformat{\section}
  {\normalfont\fontsize{14}{0}\bfseries}{\thesection}{0.5em}{}
\begin{document}
 \thispagestyle{empty} 
\raggedright
\LARGE
Astro2020 Science White Paper \vspace{0.25cm}

\LARGE
\textbf{Achieving Transformative Understanding of Extreme \\  Stellar Explosions with ELT-enabled\\ Late-time Spectroscopy} \linebreak
\normalsize

\noindent \textbf{Thematic Areas:} \hspace*{60pt} $\square$ Planetary Systems \hspace*{10pt} $\square$ Star and Planet Formation \hspace*{20pt}\linebreak
$\boxtimes$ Formation and Evolution of Compact Objects \hspace*{31pt} $\boxtimes$ Cosmology and Fundamental Physics \linebreak
  $\boxtimes$  Stars and Stellar Evolution \hspace*{1pt} $\square$ Resolved Stellar Populations and their Environments \hspace*{40pt} \linebreak
  $\square$    Galaxy Evolution   \hspace*{45pt}
   $\boxtimes$            Multi-Messenger Astronomy and Astrophysics \hspace*{65pt} \linebreak
  
\textbf{Principal Author:}

Name:	Dan Milisavljevic
 \linebreak						
Institution:  Purdue University, Department of Physics and Astronomy
\linebreak
Email: \href{mailto:dmilisav@purdue.edu}{dmilisav@purdue.edu}
 \linebreak

\textbf{Co-authors:} R.\ Margutti (Northwestern), R.\ Chornock (Ohio U), A.\ Rest (STScI), M.\ Graham (U Washington), D.\ DePoy (Texas A\&M), J.\ Marshall (Texas A\&M), V.\ Z.\ Golkhou (U Washington), G.\ Williams (MMTO), J.\ Rho (SETI), R.\ Street (LCO), W.\ Skidmore (TMTIO), Y.\ Haojing (U Missouri-Columbia), J.\ Bloom (UC Berkeley),  S.\ Starrfield (Arizona State), C.-H.\ Lee (NOAO), P.\ S.\ Cowperthwaite (Carnegie), G.\ Stringfellow (U Colorado), D.\ Coppejans (Northwestern), G.\ Terreran (Northwestern), N.\ Sravan (Purdue),  O.\ Fox (STScI), J.\ Mauerhan (Aerospace Corporation), K.\ S.\ Long (STScI/Eureka), W.\ P.\ Blair (Johns Hopkins), P.\ F.\ Winkler (Middlebury), M.\ R.\ Drout (U Toronto), J.\ Andrews (U Arizona), W.\ Kerzendorf (NYU/CCA), J.\ C.\ Wheeler (UT Austin), A.\ V.\ Filippenko (UC Berkeley), N.\ Smith (U Arizona), B.\ D.\ Metzger (Columbia), M.\ Modjaz (NYU), R.\ A.\ Fesen (Dartmouth), E.\ Berger (Harvard), P.\ Garnavich (Notre Dame), R.\ A.\ Chevalier (U Virginia)
  \linebreak

\textbf{Abstract  (optional):}

Supernovae are among the most powerful and influential explosions in the universe. They are also ideal multi-messenger laboratories to study extreme astrophysics. However, many fundamental properties of supernovae related to their diverse progenitor systems and explosion mechanisms remain poorly constrained. Here we outline how late-time spectroscopic observations obtained during the nebular phase (several months to years after explosion), made possible with the next generation of Extremely Large Telescopes, will facilitate transformational science opportunities and rapidly accelerate the community towards our goal of achieving a complete understanding of supernova explosions. We highlight specific examples of how complementary GMT and TMT instrumentation will enable high fidelity spectroscopy from which the line profiles and luminosities of elements tracing mass loss and ejecta can be used to extract kinematic and chemical information with unprecedented detail, for hundreds of objects. This will provide uniquely powerful constraints on the evolutionary phases stars may experience approaching a supernova explosion; the subsequent explosion dynamics; their nucleosynthesis yields; and the formation of compact objects that may act as central engines.

\pagebreak
\pagenumbering{arabic} 

\section{Time Domain Astronomy as a Driver of Scientific Discovery}

The expanding zoo of astronomical transients discovered in all regions of luminosity vs.\ duration phase space has become one of the most important driving forces of scientific discovery in extreme physics and astronomy. Increasingly sophisticated all-sky surveys coupled with dramatic improvements in coordination between multi-wavelength (gamma-ray through radio) and multi-messenger (gravitational wave and neutrino) facilities have enabled a fuller characterization of transients that reflect the diverse consequences of stellar evolution in the forms of disruptions, eruptions, and explosions. Consequently, many transients have forced radical revisions to long accepted models of stellar evolution with wide-reaching influence. These discoveries, which will increase by orders of magnitude in the upcoming decades, are shaping the priorities of the next generation of transformative science facilities such as the Large Synoptic Survey Telescope (LSST) and next-generation Extremely Large Telescopes (ELTs).\footnote{See Astro2020 Science White Papers ``Multi-Messenger Astronomy with Extremely Large Telescopes'' (Chornock et al.\ 2019) and ``Discovery Frontiers of Explosive Transients: An ELT \& LSST Perspective'' (Graham et al.\ 2019).}

\section{Unresolved Issues Understanding Supernovae and Their Consequences}

Supernovae (SNe) are among the most powerful and influential transients in the universe. They affect the energy balance, global structure, and chemical make-up of galaxies \citep{DS08,Powell11}; are a primary source of nucleosynthesis \citep{Wiersma11,Nomoto13} and a major source of dust \citep{Todini01}; produce neutron stars, black holes, and some gamma-ray bursts \citep{Woosley02}; and are accelerators of cosmic rays \citep{Helder09}. Core-collapse SNe (Type II/Ib,c) are also prodigious emitters of neutrinos \citep{Hirata87} and are likely strong producers of Galactic gravitational waves \citep{Murphy09,Andresen17} that can now be studied with advanced multi-messenger facilities \citep{Abbott16,Abbott17}. Thermonuclear SNe (Type Ia) associated with white dwarf explosions are used as standard candles in determining cosmological distances and played an important role in the discovery that the universe's expansion is accelerating \citep{Riess98,Perlmutter99}.

\smallskip

Many milestones of theoretical and observational understanding of SNe have been achieved in the past 10-15 years. These achievements include detection and deep limits on progenitor stars and surviving companions from high-resolution imaging  \citep{Smartt09,Li11,Schaefer12,Eldridge13}; explosion asymmetries measured via spectropolarimetry \citep{Patat17}; light echo spectroscopy \citep{Krause08,Rest11}; advances in understanding the role between mass loss and progenitor structure \citep{SA14,Smith14ARAA}; 3D simulations that have made tremendous progress towards successful core-collapse \citep{Muller16} and thermonuclear \citep{RS18} explosions; and links forged between unresolved extragalactic SNe and resolved Galactic SN remnants (SNRs) \citep{MF16-handbook}. \textbf{However, despite these advances, many key open questions regarding the fundamental properties of SNe remain outstanding} (see, e.g., \citealt{Janka12}, \citealt{Maguire17}, and \citealt{Burrows18}). Considering the wide-ranging impacts that SNe have in many fields of astrophysics, it is concerning that presently no robust mapping exists between SN classifications, their many progenitor systems, and their potential explosion mechanisms.

\begin{figure}[ht]
\centering
\includegraphics[width=0.95\linewidth]{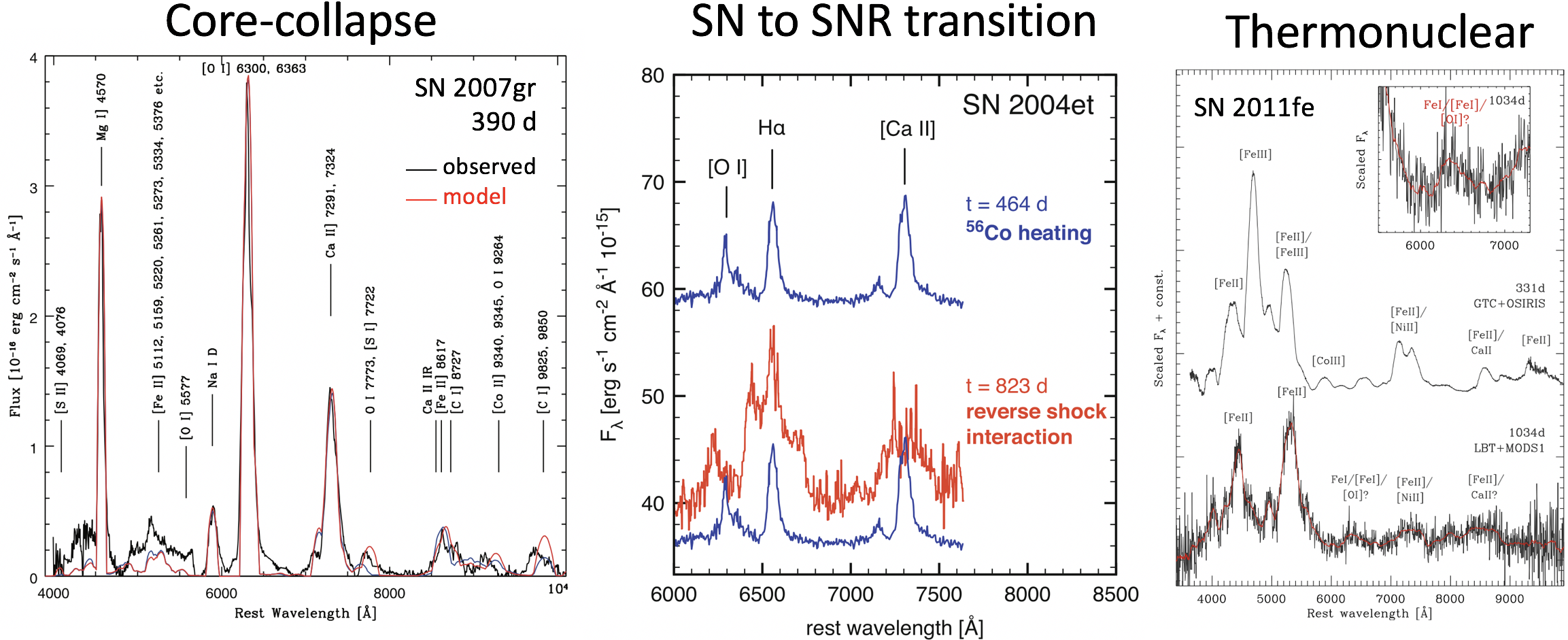}
\caption{Examples of late-time spectroscopy. {\it Left}: Day 390 observed spectrum of SN\,2007gr compared to models presented in \citet{Mazzali10} that can estimate explosion parameters.  {\it Middle}: Two epochs of SN\,2004et illustrating the rarely observed transition from SN to SNR \citep{MF16-handbook}. {\it Right}:  Extremely late 1034 day spectrum of Type Ia SN\,2011fe compared to a spectrum $\approx 1$ yr after explosion \citep{Taubenberger15}. The prominent [Fe~III] emission feature around 4700~\AA\ has entirely faded away and other features are collectively redshifted. There is no confident interpretation for this evolution. \textbf{ELTs will enable dramatic improvements to the quality of data and the number of objects that can be studied in this way, eliminating present bias towards nearby and/or interacting SNe.}}
\label{fig:four}
\end{figure}

\section{Transformative Late-time Spectroscopy}

A key observational tool that has enabled significant advancement in our understanding of SNe has been  {\it late-time spectroscopy} (Figure 1). Analyses of emission-line profiles of ejecta-tracing elements during the ``nebular phase'' beginning several months after outburst probe the chemical and kinematic properties of the metal-rich ejecta, thereby yielding clues about explosion dynamics and geometry. \textbf{The power of late-time observations comes from their ability to simultaneously probe inner regions of the debris cloud that are normally hidden when the SN is brightest, and nearby environments sculpted by the progenitor system's mass loss in the terminal stages that precede an explosion.} Models of emission have increased in sophistication and are being used to estimate progenitor mass and explosion properties \citep{Blondin12,Mazzali15,Jerkstrand17}. Asymmetry in emission line profiles has been linked to aspherical explosions and large-scale clumping in SN ejecta \citep{Maeda08,Modjaz08}. For Type Ia explosions, late-time spectroscopy has been used to investigate off-center ignition \citep{2010Natur.466...82M} vs.\ binary collision \citep{2009ApJ...705L.128R}, nucleosynthesis and geometry effects, and signatures of a non-degenerate companion star \citep{2015MNRAS.454.1948G,Boty18} in unprecedented detail \citep{Kozma05}.

\smallskip

\textbf{Late-time spectroscopy has revolutionized our understanding of the eruptions and explosions massive stars may experience in the months to years preceding core collapse} \citep{Foley07,Pastorello07,Ofek14,Andrews18}. Although brief timescales between eruption and SN explosion have been anticipated in special cases of very massive stars \citep{Woosley07,QS12}, a growing number of systems are being discovered that fall outside most theoretical regimes and that challenge many long held notions of stellar evolution, including mass loss through steady winds (Figure 2; \citealt{Margutti14}; \citealt{SA14}; \citealt{Milisavljevic15}; \citealt{Mauerhan18}).  Over the past few years, additional systems have been found that extend the timescales that core instabilities may dynamically affect stellar envelopes, and demonstrate that precursor activity is not limited to H-rich systems. \textbf{The predictions of massive-star evolution models are used in many areas of astronomy, and thus the current disconnect between theory and observation presents a real problem.} Specifically, uncertainties in mass loss estimates affect predictions for ionizing radiation and wind feedback from stellar populations, as well as conclusions about star formation rates and initial mass functions in external galaxies \citep{Smith14ARAA}. \textbf{Fully characterizing precursor eruptions in SN progenitor systems may be a critical ingredient to understanding the core-collapse process} \citep{AM11}.

\begin{figure}
\centering
\includegraphics[width=0.95\linewidth]{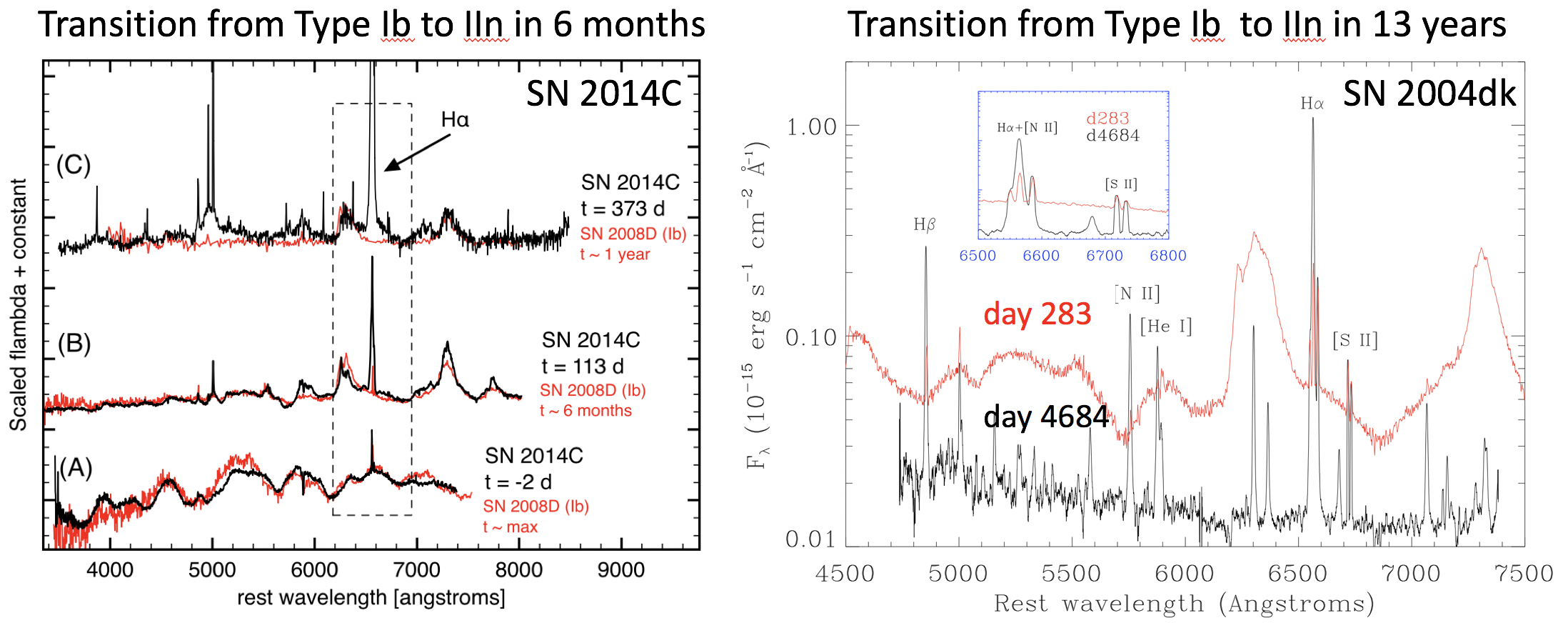}

\caption{Late-time spectroscopy probes progenitor star mass loss and evolutionary changes approaching core collapse. {\it Left}: SN 2014C slowly transformed from a hydrogen-poor Type Ib explosion to a hydrogen-rich strongly interacting Type IIn, which is consistent with a delayed interaction between an H-poor progenitor star's SN explosion and a massive H-rich envelope that had been ejected approximately 500 years before core collapse \citep{MM18}. {\it Right}: Deep late-time optical spectroscopy of the Type Ib explosion SN2004dk 4684 days (13 years) after discovery revealed prominent intermediate-width H$\alpha$ emission consistent with a short-lived Wolf-Rayet phase that followed a slow dense wind that persisted for millennia \citep{Mauerhan18}. \textbf{These late-time interaction examples challenge many long held notions of stellar evolution, especially mass loss through steady winds.}}

\label{fig:two}
\end{figure}

\section{The Need for ELTs}

Presently, late-time spectroscopy is biased towards epochs $< 400$ days after the explosion. \textbf{Beyond two years, our knowledge of the spectroscopic evolution of SNe is extremely limited.}  This introduces an urgent need for a US-ELT Program. Following SNe through the SN-to-SN remnant (SNR) transition via late-time observations has enabled many exciting fundamental connections to be made between SNe, their progenitor stars, their products (both compact and ejected debris, including dust), and resolved SNR analogs \citep{Fesen99,Larsson13,MF15,Bevan17,Long19,Rho19}. 

\smallskip

\textbf{The primary challenge of obtaining late-time spectroscopy on a statistically significant sample (hundreds of objects) is simply light gathering power: the majority of SNe fade at least eight magnitudes below peak brightness in their first 2 years (often even more rapidly), limiting observations to 1 year or so after maximum light when they are at apparent magnitudes $< 22$.}  Thus, the large aperture of an ELT in combination with the high efficiency of its instruments both in the optical and NIR can revolutionize late-time investigations of SNe both in the volume of space that can be sampled (and hence the number of objects to study) and the quality of data to be obtained.  TMT and GMT are both required to cover northern and southern hemispheres and provide additional observing redundancy in overlapping regions of the sky. Spectroscopic observations generally require moderate resolutions ($R \sim 2000$) over the broadest possible wavelength range (0.32 -- 2.4 micron). GMT+GMACS, TMT+WFOS, and TMT+NFIRAOS+IRIS are potentially the best combinations. However, new exciting opportunities for important complementary science will be made possible with higher resolution configurations available with TMT+WFOS, HROS, and IRIS, and GMT+G-CLEF and GMTIFS. We note that \textbf{NIR wavelengths enable powerful line diagnostics, avoid extinction, and are particularly sensitive to dust formation, but presently late-time spectroscopy at these wavelengths is extremely limited. New ELT instruments like GMTIFS and IRIS will make NIR late-time spectroscopy much more common.} Furthermore, \textbf{the growing awareness that asymmetry plays a significant role in SN explosions provides strong motivation for spectropolarimetry, which is often the only direct measure of key characteristics including intrinsically non-spherical material and strong magnetic fields.} Spectropolarimetry is rarely applied ($ <200$ SNe, and most for a single epoch) because of the challenging nature of the observations. ELTs have the potential to easily overcome these challenges, and it is critical that GMT + TMT instrumentation designs not
preclude this capability.

\smallskip

\textbf{Late-time spectroscopy of increased resolution and depth for hundreds of SNe will enable line diagnostics that can robustly distinguish between the various mechanisms of late-time emission that, in turn, tightly constrain progenitor systems and explosion mechanisms.} A key diagnostic is in the observed changes in velocity widths of emission line profiles. In circumstellar interaction scenarios where the reverse shock penetrates into deeper layers of ejecta, the velocity widths of emission lines are expected to narrow with time.  Alternatively, in scenarios involving a pulsar wind nebula where emission is powered by photoionization, line widths are anticipated to remain constant or broaden because of acceleration by the pulsar wind bubble.  Some SNe show signs of hosting pulsar wind nebulae or black holes at X-ray wavelengths \citep{Patnaude11,Dittmann14}, but data of sufficient signal-to-noise do not exist to confirm this at optical wavelengths.  Notably, this emission line diagnostic will be a key test of the hypothesized connections between long-duration gamma-ray bursts, superluminous SNe (SLSNe), more ``ordinary'' SNe, and potentially even Fast Radio Bursts \citep{Milisavljevic13,Nicholl16,Metzger17}.  The ELTs will enable the late-time precision spectroscopy required to detect broadening in optical and NIR emission line widths at the rate of 1-5\%\,yr$^{-1}$  in objects with apparent magnitudes  m$_R > 24$ mag (Figure 3; \citealt{Milisavljevic18}).

\smallskip

\textbf{It is highly anticipated that late-time precision spectroscopy will probe uncharted frontiers of the diverse array of Type Ia explosions and related phenomena, including Type Iax \citep{Foley16} and Ca-rich transients} \citep{Kasliwal12,Milisavljevic17,De18}.  The only available spectrum of a Type Ia SNe more than 1000 d after its explosion revealed a dramatic transformation with unexplainable features \citep{Taubenberger15,Black17}. An ELT will increase the number of objects from 1 to $\sim 50$ in less than five years of operation.  Generally, the number of core-collapse SNe observed at late epochs is much larger than SNe Ia, and this is thought to be a reflection of the immediate environment: while SNe Ia tend to be in low-density environments, core collapse SNe explode in higher density circumstellar material (CSM) left behind by mass loss of the massive star progenitor system. However, this may be an observational bias towards nearby objects observed in the first year after explosion. \citet{Graham19} observed late-onset interaction between a SN\,1991T-like overluminous explosion and surrounding CSM $\sim 600$ days after its light-curve peak (see also \citealt{Bochenek18}). ELTs will have a vital role in constraining the extent to which other Type Ia SNe experience similar rebrightenings and the composition of the mass loss material.

\medskip

\textbf{In summary, ELT-enabled late-time precision spectroscopy will facilitate transformational understanding of supernovae and their progenitor systems.} The development of 30-m class telescopes and instrumentation is a critical requirement to fully investigate the evolutionary phases stars may experience approaching a supernova explosion; the subsequent explosion dynamics; and the formation of compact objects that may act as central engines. Any attempt to model and correctly interpret the electromagnetic+gravitational wave+neutrino signals of supernovae will require solid understanding of these fundamental properties.

\begin{figure}
\centering
\includegraphics[width=0.52\linewidth]{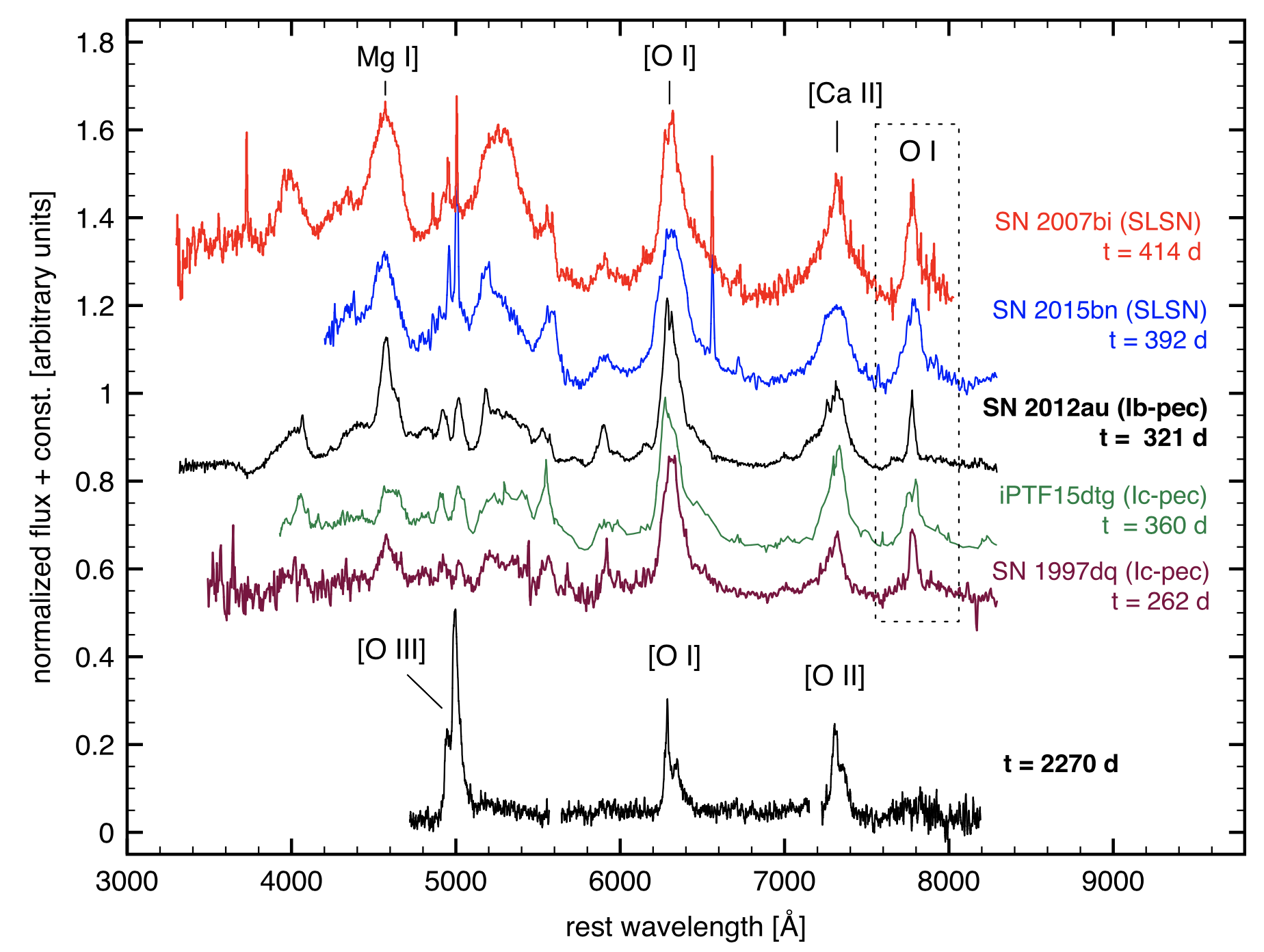}
\includegraphics[width=0.3\linewidth]{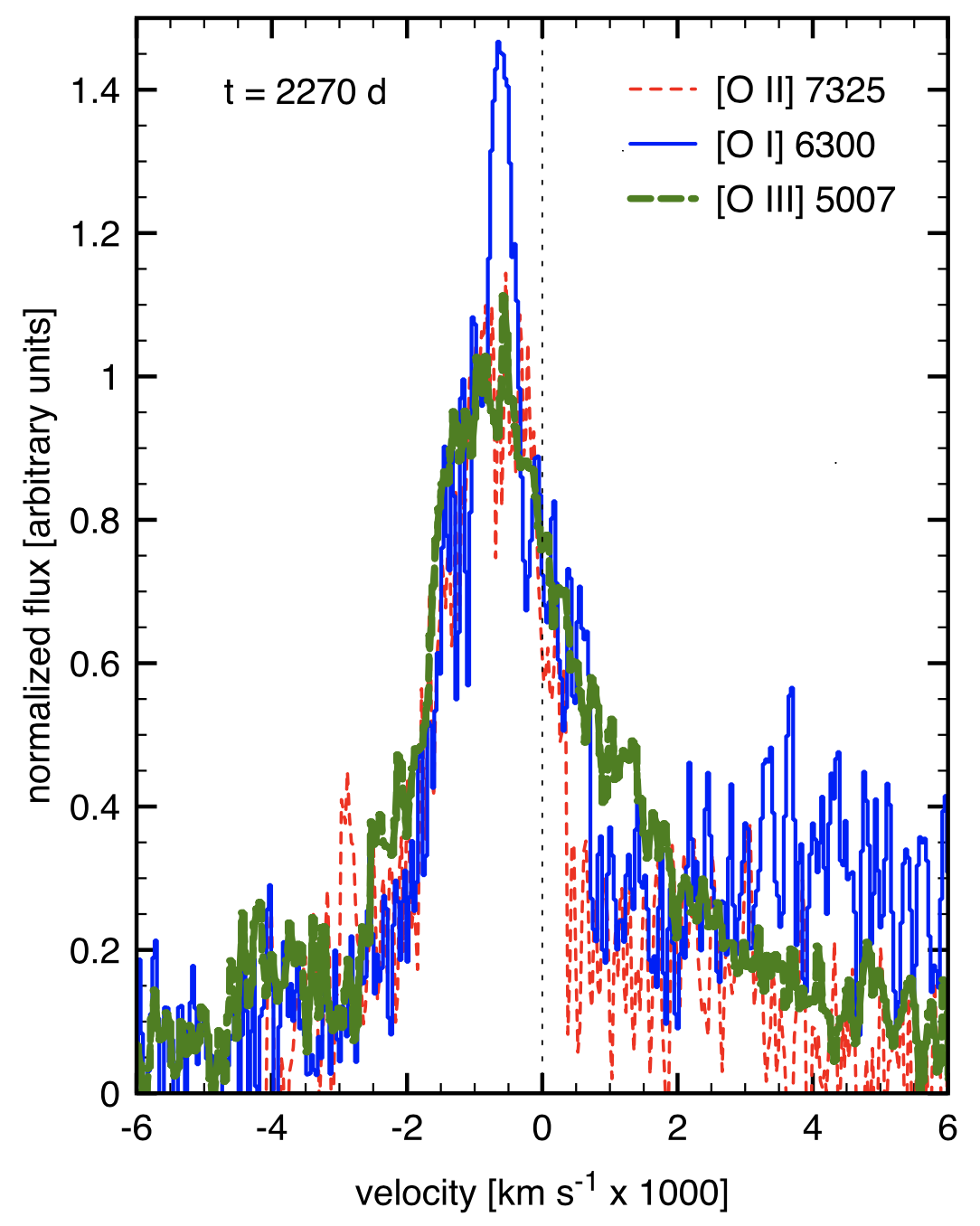}
\caption{{\it Left}: Spectra of the H-poor SLSN 2015bn over 400 days post-explosion are identical to a number of energetic SNe Ic and the SN Ib 2012au indicating similar core conditions \citep{Nicholl16}. Continued monitoring of SN\,2012au to $> 6$~yr post-explosion showed a dramatic change in the emissions consistent with a newly-formed pulsar wind nebula (PWN) exciting O-rich ejecta. SLSNe may undergo the same transformation at extremely late epochs \citep{Milisavljevic18}. {\it Right}: Enlarged emission line profiles of SN\,2012au. \textbf{ELTs will make it possible to observe SLSNe at these epochs and test predictions that emission line widths should remain constant or increase with time if influenced by pulsar/magnetar wind nebulae.}}

\label{fig:three}
\end{figure}

\pagebreak

\bibliographystyle{aasjournal_DD}

\end{document}